# Resolving the controversy on the glass transition temperature of water?

S. Capaccioli[a,b*], K. L. Ngai[a,c*]

[a]*CNR-IPCF, Dipartimento di Fisica, Largo Bruno Pontecorvo 3,I-56127, Pisa, Italy*
[b]*Dipartimento di Fisica, Università di Pisa, Largo Bruno Pontecorvo 3 ,I-56127, Pisa, Italy*
[c]*State Key Lab of Metastable Materials Science and Technology, and College of Materials Science and Engineering, Yanshan University, Qinhuangdao, Hebei, 066004 China*

___________________________________________________________________________

**Abstract**

We consider experimental data on the dynamics of water (1) in glass-forming aqueous mixtures with glass transition temperature $T_g$ approaching the putative $T_g$=136 K of water from above and below, (2) in confined spaces of nanometer in size and (3) in the bulk at temperatures above the homogeneous nucleation temperature. Altogether, the considered relaxation times from the data range nearly over 15 decades from $10^{-12}$ to $10^3$ s. Assisted by the various features in the isothermal spectra and theoretical interpretation, these considerations enable us to conclude that relaxation of un-crystallized water is highly non-cooperative. The exponent $\beta_K$ of its Kohlrausch stretched exponential correlation function is not far from having the value of one, and hence the deviation from exponential time decay is slight. Albeit the temperature dependence of its α-relaxation time being non-Arrhenius, the corresponding $T_g$-scaled temperature dependence has small steepness index $m$, likely less than 44 at $T_g$, and hence water is not 'fragile' as a glassformer. The separation in time scale of the α- and the β-relaxations is small at $T_g$, becomes smaller at higher temperatures, and they merge together shortly above $T_g$. From all these properties and by inference, water is highly non-cooperative as a glass-former, it has short cooperative length-scale, and possibly minimal configurational entropy and small change of heat capacity at $T_g$ compared with other organic glass-formers. This conclusion is perhaps unsurprising because water is the smallest molecule. Our deductions from the data rule out that the $T_g$ of water is higher than 160 K, and suggest that it is close to the traditional value of 136 K.

---

[*] capacci@df.unipi.it  Tel. +39-0502214537, Fax. +39-0502214333
[*]Ngai@df.unipi.it, Tel. +39-0502214322, Fax. +39-0502214333

# 1. Introduction

There is a long history in the research on the relaxation and the glass transition of amorphous water [1]. The generally accepted view is that the onset glass transition $T_g$ for vitreous water occurs at about 136 K for amorphous solid water (ASW) and hyperquenched glassy water (HGW). ASW was discovered by Burton and Oliver [2] in deposits of water vapors on a substrate kept at a temperature of ~77 K. HGW was produced by Mayer as millimeter thick deposits on a copper disc kept at a temperature of ~77 K. [3, 4]. The $T_g \approx 136$ K was reported for both ASW and HGW on the base of calorimetric determination by Johari, Hallbrucker and Mayer in 1987 [5], and dielectric relaxation measurement of annealed ASW films by scanning temperature at a fixed frequency of 1 kHz by the same workers [6]. This value also was supported by linear extrapolation of data of $T_g$ from glass-forming aqueous mixtures [7, 8]. However, since 2002, Angell and coworkers [8, 9] had raised question on the validity of $T_g \approx 136$ K for water by suggesting that the feature observed by differential scanning calorimetry (DSC) of ASW is not the $T_g$ endotherm but a "shadow-$T_g$ peak" or "sub-$T_g$ peak" appearing in the $0.7T_g$ to $0.8T_g$ range in the DSC scan of some glasses pre-annealed at a specific temperature [10, 11, 12, 13, 14, 15, 16, 17, 18]. From this, Angell proposed that $T_g$ of water is between 165 and 180 K, and is unobservable [8]. In response to this, Johari provided several criteria to distinguish sub-$T_g$ peak from $T_g$ endotherm by their different DSC features, and concluded that the well-known endothermic feature observed for glassy water cannot be regarded as a sub-$T_g$ peak, and it is a $T_g$ endotherm [19]. In the following years, Angell and co-workers [20, 21, 22] again suggested $T_g$ of water must lie above 160 K. This time the argument was based on the difference of the measured dielectric tanδ between ASW [20] and glassforming water-like aqueous solutions, $H_2O$-$H_2O_2$ and $H_2O$-$N_2H_4$. Johari had reexamined the dielectric tanδ data of ASW and demonstrated that water is a liquid over the 136–155 K range, thus negating the suggestion of revising the $T_g$ of water to lie above 160 K [23]. The controversy on the location of $T_g$ of water continued with the latest view of Angell [24] that the glass transition in water is an order-disorder transition that occurs in the range from 150 to 250 K, and the criticism on its premise by Johari et al. [25].

Despite the difference in the interpretation of DSC and dielectric relaxation data of ASW, both sides have confined their consideration of the experimental data that are directly related to the question of the location of $T_g$. In this work we take a different route to locate the $T_g$ of water by exploiting altogether various dynamic properties of water in the bulk state, in nano-confinement, and in mixtures with hydrophilic solutes by various experimental techniques over broad range of time scales.. This cannot be done with ASW because of various complications including ASW and HGW are known to crystallize rapidly at $T$>140 K [26, 27]. Instead we study the glass-forming aqueous mixtures with large molar fraction of water having $T_g$ approaching the putative $T_g$=136 K from above and below. The use of aqueous mixtures to deduce the $T_g$ of pure water has been done before by Angell and Sare [28] by extrapolation of binary solution $T_g$ data on many salt + water systems and aqueous molecular liquids to pure water. The extrapolations yield $T_g(H_2O)$=139 K, as shown in a summarizing figure by MacFarlane and Angell [7]. We differ from the previous works by considering the dynamic properties of various aqueous mixtures and bulk water when combined over about 15 decades of frequency. By this approach, we have at our disposal several ways to deduce the dynamic properties of water, which are used in turn to locate its $T_g$.

In glass-forming aqueous mixtures, water participates in the structural α-relaxation as well as manifests itself as the secondary β-relaxation, which can rotate and translate independently after hydrogen bonds have been broken (for details on this issue see Refs.[29, 30]). As established before for ordinary glass-formers and binary mixtures, the separation between the α- and β-relaxations at $T_g$ of the aqueous mixtures is used as an indicator of the degree of cooperativity or length scale of the α-relaxation. We observe decrease in separation of the α- and β-relaxations at $T_g$ of on decreasing $T_g$ of aqueous mixtures towards the putative $T_g$=136 K of water from above. The separation becomes sufficiently small to enable us to conclude the degree of cooperativity or length scale of the α-relaxation of the mixtures is small, and *a fortiori* the same conclusion applies to water because of the expected enhanced cooperativity of water in the presence of less mobile solute in these aqueous mixtures. This deduction is supported by the observation of the properties of the relaxation of water confined in nanometer spaces where cooperativity is totally suppressed due to reduction of length-scale. The relaxation found is almost identical to the β-relaxation of water in the aqueous mixtures with high water content. Mixture of 1-propanol with water has $T_g$ significantly lower than 136 K [31, 32]. Water is slower than 1-propanol, in the mixture, and the dynamics of the water component observed have the signature of the α-relaxation. The β-relaxation of the water component is either absent or cannot be distinguished from the α-relaxation. This is in contrast to other aqueous mixtures with $T_g \geq 136$ K where water is the faster component and it participates in the α-relaxation in concert with the solute as well as principally responsible for the well resolved β-relaxation. We consider this drastic change of role of water in mixture with 1-propanol as another indication of low degree of cooperativity of water, which by virtue of the Arrhenius temperature dependence of its relaxation time appears to have been totally removed by mixing with 55 mole% of the more mobile 1-propanol. The α-relaxation time $\tau_\alpha$ of the water component in the mixture with 1-propanol reaches 100 s at the temperature of $T_g \sim 110$ K. $\tau_\alpha$ of neat 1-propanol also reaches 100 s at $T_g \approx 100$ K. These values of $T_g$ suggests that $T_g$ of water may not lie above 160 K because it is hard to rationalize the large reduction of $T_g$ of water by more than 50 K to be 10 degrees above $T_g \approx 100$ K of neat 1-propanol by mixing with 55 mole% of it. More features of the dynamics of water in mixture and in confinement are brought out in the following sections to support our conclusions that water is highly non-cooperative, possibly with little configurational entropy and small change of heat capacity at $T_g$, which is likely no larger than 136 K. We present new data on the dynamics of mixtures of ethylene glycol (EG) and water at different concentrations as well as new analysis of results on other aqueous systems, mixed or confined, from literature. It has to be pointed out that numerous papers in literature have also used study of aqueous solutions of strongly hydrogen-bonding liquids and electrolytes to infer the $T_g$ of water [7, 20, 21, 28, 33]. Properties of mixed, confined and surface water could be different from that of bulk water due to the modification of hydrogen bond network and the local arrangement. Notwithstanding, this does not rule out the applicability of locating the water molecules dynamics within the upper and lower bounds obtained from these other systems used for comparison.

2. Experiment

We studied mixtures of ethylene glycol (EG) and water at two different concentrations: EG:H$_2$O weight proportion of 60:40 and 50:50 (respectively 30.3% and 22.5% molar fraction of EG). Ethylene glycol (anhydrous, 99.8%) was purchased from Aldrich. Distilled and deionized water (electric conductivity lower than 18.3 μS m$^{-1}$) was prepared by an ultra-pure water distiller (Millipore, MILLI-Q Lab). Both mixing of components and loading of the sample holder were performed in controlled atmosphere (dry nitrogen).
Dielectric spectroscopy measurements on EG:H$_2$O mixtures were done in the frequency range from 10 mHz to 10 GHz, by means of a combination of the Novocontrol Dielectric Analyser and the Agilent Network Analysers 8722D and 8753ES. Dielectric spectra, acquired isothermally at different temperatures in subsequent steps after appropriate thermal equilibration, were obtained according to two protocols in two temperature range: (i) after a fast quenching (more than 40 K/min), on heating from the glassy (100 K) to the liquid state (200 K); (ii) slowly cooling from room temperature down to 200 K. Protocol (i) was used for low frequency measurements (up to 10 MHz), (ii) for high frequency measurements. Further details on the experimental technique can be found in ref.[34].

## 3. Results and Discussion: Support from dielectric relaxation data of aqueous mixtures

*3.1 Aqueous mixtures of oligomers of ethylene glycol and ethylene glycol*

If the aqueous mixture has large enough molar fraction of water in, the α-relaxation necessarily involves water in cooperative motion with the solute. The Coupling Model (CM) [35, 36, 37, 38, 39] can be used to determine the coupling parameter $n$ of the water component in the mixtures, which is an indicator of the degree of cooperativity of the α-relaxation involving the water in the mixture. This is carried out by using

$$\tau_\alpha = \left[t_C^{-n}\tau_0\right]^{1/(1-n)} \approx \left[t_C^{-n}\tau_\beta\right]^{1/(1-n)}. \tag{1}$$

The first part of eq.(1) comes from the exact CM relation between the α-relaxation time $\tau_\alpha$ and the primitive relaxation time $\tau_0$ of the CM. The second part originates from the approximate relation, $\tau_0 \approx \tau_\beta$, leading in turn to the approximate relation between $\tau_\alpha$ and the observed JG β-relaxation time $\tau_{JG}$ of water in the mixture. Taking $t_c$=2 ps as the same for other organic glassformers, $n$ can be obtained from eq.(1) now rewritten as

$$n \approx (\log\tau_\alpha - \log\tau_\beta)/(\log\tau_\alpha + 11.7), \tag{2}$$

Furthermore, on changing the solute systematically in such higher water content mixtures and observing the trend of the corresponding change of $n$, some insight into the dynamics of bulk water possibly may be obtained. This can be done by utilizing the 35 to 40 wt.% water mixtures with the oligomers of ethylene glycol. Shown before in Fig.6 of Ref.[30] are the $T$-dependences of the α- and water β-relaxation times of the 35 wt.% water mixtures with tri-ethylene glycol (3EG) [40] and with di-ethylene glycol (2EG) [41], and 40 wt.% of water mixture with ethylene glycol (EG) [42]. From $\tau_\alpha$ and $\tau_\beta$ available at the lowest temperature for each mixture, the values of coupling parameter $n$ calculated by eq.(2) are 0.24, 0.19, and 0.10 for mixtures with 3EG, 2EG, and EG respectively. These values of $n$ are plotted against the number of carbon atoms $N_C$ of the solute in the inset of Fig.6 of Ref.[30]. A linear extrapolation of the data as shown suggests that $n$ is nearly zero when there is no more carbon atoms. This extrapolated value of $n$ cannot be strictly identified with that for bulk water. Nevertheless, with or without extrapolation, the $n$ vs. $N_C$ plot indicates that $n$ of bulk water at low temperatures, say below the dielectric $T_g(\tau_\alpha=100\text{ s}) \approx 162$ K, 154 K, and 137.8 K of 35 wt.% water-3EG, 35 wt.% water-2EG, and 40 wt.% water-EG respectively, is small. Since continued addition of water to 3EG, 2EG or EG monotonically lowers the dielectric $T_g$ from that of the neat solutes to a lower value [31, 30], for example from 178 K of pure 3EG to 162 K with addition of 35 wt.% of water, it is clear that water is the faster component in all these mixtures. Conversely water is slowed down by the less mobile solutes, which immediately implies that the glass transition temperature $T_g$ of pure water has to be less than the observed $T_g$ of these aqueous mixtures, the lowest one is 137.8 K for 40 wt.% water-EG (~144 K by calorimetry versus 155 K of neat EG [33]). This deduction supports that the dielectric (calorimetric) $T_g$ of pure water cannot be above 137.8 K (144 K). On the other hand, it is hard to rationalize the observed shift of the dielectric $T_g$ down to 137.8 K by mixing EG with 40 wt.% of water if the $T_g$ of pure water is 165 K or higher [43, 44, 20, 21, 22, 24]. This is because the dielectric $T_g$ of neat EG located at about 150 K [33] is even lower than the hypothesized $T_g$ =165 K for water.

The steepness or "fragility" index $m$ is 55 for 35 wt.% water-2EG and 44.2 for 40 wt.% water-EG. From the monotonic decrease of $m$ with the number of carbon atoms $N_C$ of the solute, we deduce that $m$ of pure water will be smaller. There is usually a correlation between $m$ and $n$ for glassformers belonging to the same class [45, 35]. Thus, the expected small $m$ of pure water is consistent with the small $n$ deduced in Fig.6 of Ref.[30] from a different dynamic property of the mixtures, altogether indicating that water has low level of cooperativity as a glassformer.

*3.2 Aqueous solutions of $H_2O$-$H_2O_2$ and $H_2O$-$N_2H_4$*

The dielectric relaxation data of aqueous solutions of two water-like molecules, hydrazine $N_2H_4$ and hydrogen peroxide $H_2O_2$, in the neighborhood of their glass transition temperatures, $T_g$ offer another opportunity to deduce the dynamics of pure water and its value of $T_g$. By extrapolation of $T_g$ data of mixtures of $N_2H_4$ with various acids, Sutter and Angell [46] determined $T_g$ of neat $N_2H_2$ to be 125 K, which they considered to be reliable to ±5 K. Similarly for neat $H_2O_2$, the value for $T_g$ is 139 K. Isothermal and isochronal dielectric relaxation measurements of supercooled aqueous solutions of hydrazine with compositions 20, 26.5, and 33 mol % $N_2H_4$ and aqueous solution of 35 mol % $H_2O_2$ were made by Minagouchi et al. at temperatures above and below $T_g$ [20, 21]. The α-relaxation of the hydrazine solutions has narrow frequency dispersion at all temperatures, narrower than seen before in other aqueous mixtures, and corresponding to correlation function not far from being exponential function of time. No secondary relaxation including the JG β-relaxation of the water component has been resolved. This behavior contrasts to other aqueous mixtures with comparable mol % of water, where the JG β-relaxation of water is well separated from the α-relaxation and hence well resolved. By courtesy of Ranko Richert, the isothermal dielectric loss spectra of solution of 26.5 mol % $N_2H_4$ at temperatures above $T_g$ are reproduced in a log-log plot in Fig.1. The fits by the one-sided Fourier transform of the Kohlrausch function,

$$\phi(t) = \exp\left[-\left(\frac{t}{\tau_\alpha}\right)^{1-n}\right] \qquad (3)$$

at three temperatures are shown by the dashed lines with $n$=0.16, 0.16, and 0.19 (from right to left, at temperatures of 154 K, 150 K, and 146 K). The arrows indicate the location of the primitive frequency, $f_0 \equiv 1/(2\pi\tau_0)$, where $\tau_0$ is calculated by eq.(1) from $f_\alpha \equiv 1/(2\pi\tau_\alpha)$, $n$, and $t_c$=2 ps for each temperature. The close proximity of $f_0$ to $f_\alpha$ explains why the JG β-relaxation of water is not resolved in the solution of 26.5 mol % $N_2H_4$. The same situation holds for the other solutions of hydrazine of 20 and 33 mol % $N_2H_4$. The width of the α-relaxation of the aqueous solution of 35 mol % $H_2O_2$ is a bit broader, yielding a value of $n \approx 0.23$. In these aqueous solutions of $N_2H_4$ and $H_2O_2$ where the $n$ parameters are small, the most probable frequencies of α- and JG β-relaxation are close to each other, according to eq.(1).

Consequently the JG $\beta$-relaxation cannot be resolved from the $\alpha$-relaxation, and instead it shows up as an excess loss on the high frequency flank of the dominant α-loss peak. This phenomenon, commonly referred to in the literature as the "excess wing", has been found in many neat glassformers with small $n$ parameter [47], and its location near the α-loss peak frequency is consistent with the primitive frequency, $f_0$, calculated via Eq.(1) of the CM. This is illustrated in Fig.1 for the present case of water solution with 26.5 mol % $N_2H_4$ by the arrows to indicate the location of the calculated $f_0$ at three temperatures.

The small values of $n$ or equivalently the nearly exponential α-correlation function of all the solutions of $N_2H_4$ and $H_2O_2$ are direct indications of the highly non-cooperative nature of the α-relaxation. Usually concentration fluctuations in binary mixtures contribute to broadening of the α-relaxation if there is significant difference in mobility of the two components. The absence of broadening by concentration fluctuation suggests that the mobility or $T_g$ of water is not very different from that of $N_2H_4$. By this deduction, $T_g$ of water should not be far from $T_g$=125 ±5 K of neat $N_2H_4$ or 139 K of neat $H_2O_2$. If $T_g$ of water is 165 K or above, much higher than the $T_g$ of $N_2H_4$ or $H_2O_2$, broadening of the α-relaxation by concentration fluctuations of the two components with widely different mobility is expected, but it is not observed. Irrespective of how much concentration fluctuations contribute to the broadening of the α-relaxation of the solutions of hydrazine, the width of dispersion of the relaxation contributed by the water component must be less than the already narrow width of the solutions. By this argument and using $n$ as a measure of the degree of cooperativity, we deduce from the data of the hydrazine solutions that $n$ of water is less than 0.19, and less than 0.23 from data of the solution of $H_2O_2$.

The temperature dependence of $\tau_\alpha$ from dielectric study of Minoguchi et al. and from microwave measurements [48, 49] was fitted by the Vogel-Fulcher-Tammann-Hesse equation. The $T_g$ obtained as the temperature at which $\tau_\alpha$=100 s for the solutions with 20, 26.5, and 33 mol % $N_2H_4$ are almost identical with values of 136.3, 135.6, and 137.6 K respectively, and not far from $T_g$=125 ±5 K of neat $N_2H_4$. Moreover, the solution of 35 mol % $H_2O_2$ has $T_g$=140 K, almost the same as $T_g$=139 K of neat $H_2O_2$. The fact that $T_g$ of the solutions are insensitive to composition and have values near the $T_g$ of the neat solutes is another indication that the $T_g$ of water cannot be very different.

As an aside, the dielectric loss $\varepsilon''(f)$ data of solutions of hydrazine [20, 21] exhibits the frequency dependence, $\varepsilon''(f) \propto f^{-c}$ with $c$ positive and much less than 1, at high frequencies ($f \gg f_0 \approx f_\beta$) and temperatures below and slightly above $T_g$. This feature is the nearly constant loss (NCL) found in dielectric relaxation of many different molecular glassformers [36, 50, 51], and is interpreted as the loss from molecules while caged by the anharmonic intermolecular potential at short times before the onset of the JG β-relaxation. At a fixed frequency of 316 kHz, the temperature dependence of the NCL, $\varepsilon''(f$=316 kHz, $T)$, has been obtained for the solutions of hydrazine. Presented in Fig.2 is the result from solutions with 26.5 mol % $N_2H_4$ and 22.5 mol % (50 % wt.) of ethylene glycol (EG). There is a marked change of $T$-dependence of the NCL when crossing some temperature near $T_g$, found before in various glassformers [36, 51]. The NCL observed by dielectric relaxation is the lower frequency analogue of the mean square displacement $<u^2>$ observed by neutron scattering at shorter times. While the molecules are still caged, $<u^2(t)>$ rises slowly with power law time dependence, $\sim t^c$, where $c$ is positive and small [50, 51, 52]. This time dependence of $<u^2(t)>$ corresponds to the imaginary part of the susceptibility having the frequency dependence $\sim f^c$, which is the NCL. No genuine relaxation process occurs while the molecules are caged. Hence the loss has no characteristic time, and its power law dependence $\sim f^c$ follows naturally. This kind of behavior can be noted at high frequency in spectra at lower temperatures shown in the inset of Fig.2. For glassformers of the same type and measured in neutron scattering by the same spectrometer with the same energy resolution, there is an experimentally well established correlation between the magnitude of $<u^2(T)>$ at $T=T_g$ and non-exponentiality (or $n$) [52, 53]. For NCL observed by dielectric relaxation as in Fig.2, the comparison with other glassformers is made for all having the same α-relaxation frequency $f_\alpha$ and after normalizing the entire loss spectrum, $\varepsilon''(f)$, by the maximum loss $\varepsilon''_{max}$ of the α-loss peak spectra. This has been carried out in Fig.3 for the solution with 26.5 mol % $N_2H_4$ together with glycerol, threitol, and xylitol, with $f_\alpha=10^{-4}$ Hz. . The master curve of water solution of hydrazine at 26.5 mol % $N_2H_4$ at $T=110$ K represented by all black symbols of different shapes shown in Fig.3 was obtained by the following procedure. First, the $\varepsilon''(f)$ loss spectra at two higher temperatures, 146 K (black closed triangles, and shown before in Fig.1) and 126 K (black closed circles) were shifted horizontally to lower frequencies to superpose onto the loss data at the lower temperature of 110 K (black open circles). The height of the α-loss peak in the master curve is then normalized to unity. As shown in a log-log plot, the normalized master curve thus represents $\varepsilon''(f)$ at 110 K over the extended frequency range. The vertical arrow indicates the location of the primitive frequency, $f_0=1/(2\pi\tau_0)$, calculated from eq.(1) with $n=0.19$ determined from fit presented before in Fig.1. Similar procedure was used to construct the normalized master curve for glycerol (green symbols, at 179 K), threitol (red symbols, at 224 K), and xylitol (blue symbols, at 243 K). More details in the construction of the master curves for these polyols had been given before in Ref.[36]. The NCL shows up in each case at frequencies sufficiently high above the JG β-relaxation frequency $f_\beta$ indicated by the location of the primitive frequency, $f_0$, calculated from the known Kohlrausch exponent $(1-n)$ and $f_\alpha$ by the Coupling Model relation [35, 36]. The dashed lines with frequency dependence, $f^{-0.13}$ and $f^{-0.14}$, are used to indicate the slow variation of $\varepsilon''(f)$ at high frequencies beyond $f_0$, which is the NCL of the caged dynamics regime. In the case of xylitol, the presence of a well resolved JG β-relaxation at higher frequencies renders the NCL to appear at even higher frequencies, outside the experimental frequency range and not observed in Fig.3. Since it is a universal feature of caged dynamics, the NCL of xylitol at 243 K should be observable at higher frequencies than available in Fig.3, or within the frequency range in Fig.3 by lowering the temperature. In fact, a recent dielectric study on xylitol by Kastner et al. [54] at high frequencies in the GHz-THz range and at temperatures from 250 to 260 K has found nearly frequency independent loss over 1-2 frequency decades in the spectra, which may be interpreted as the NCL at high frequencies. The magnitude of this NCL is about two decades smaller than the maximum of the α-loss peak, which is consistent with the scenario that the NCL appears at higher frequencies than the resolved JG β-relaxation in Fig.3. Kastner et al. also found the NCL firmly established in the same frequency range as Fig.3 by lowering temperature to 150 K. The correlation between $n$ and the level of the NCL of these four glassformers with hydrogen bonds can be seen by inspection of Fig.3. Higher is $n$, bigger is the amplitude of NCL. The fact that the solution with 26.5 mol % $N_2H_4$ having the lowest level of NCL is another evidence that it has low degree of cooperativity, even lower than glycerol, and consistent with the narrow width of its α-loss peak in Fig.1.

*3.3 Aqueous mixture of 1-propanol*

The dielectric loss spectra of pure 1-propanol (PrOH) and the mixture with 20 wt.% water obtained by Sudo et al. [32, 29] are shown at 123 K in Fig.4(a) and (b) of ref.[30]. For pure PrOH, the spectra are exactly the same as that found by Hansen et al. [55] with the Debye process at lowest frequency and the unresolved α- and β-relaxations, labeled as processes I, II and III respectively in Fig.5 of Ref.[30]. For 20 wt.% water mixture, all these processes are still observed and their relaxation times are nearly the same as pure 1-propanol.

However, a new loss peak is observed at the frequency range lower than all the propanol loss peaks. This can be seen in Fig.5(b) of Ref.[30] from the loss peak labelled ν, and I′ in Fig.5(c) and (d) of Ref.[30]. This new process originates from the motion of water in the mixture, and the variation of its relaxation times with temperature can be found in Fig.6 of Ref.[30].

What distinguishes 1-propanol from all other solutes in aqueous mixtures discussed so far is that the α-relaxation time of 1-propanol, either pure or in the 20 wt.% water mixture, is shorter than even the $\tau_{JG}$ of water in the mixtures. This comes from the low $T_g$ of 1-propanol, which lies near and below 100 K. Due to this unusual circumstance, the relaxation process of water found in the mixture is slower than all processes associated with 1-propanol (see the spectrum at 123 K in Fig.5(b) in Ref.[30]). What is surprising in this case is that its properties are not like those of the JG β-process of water observed in the other aqueous mixtures discussed before. For instance, its relaxation strength, $\Delta\varepsilon_{I'}$, decreases monotonically with increasing temperature (see Fig.5(c) of Ref.[30]), which is characteristic of α-relaxation. Moreover, its dispersion width parameter $\beta_{CCI'}$ is nearly constant over a considerable range of temperature above $T_g$, which again is typical for α-relaxation, and at odds with the large increase on decreasing temperature for JG β-process. All these properties of process I′ indicate that it is the α-relaxation of water in the mixture with 1-propanol. Furthermore, the temperature dependence of $\tau_{I'}$ is nearly Arrhenius, quite unusual for α-relaxation, as shown in Fig.2 of Ref.[30]. The Arrhenius law that fits the data of $\tau_{I'}$ is given by

$$\tau_{I'}/s = 10^{-14.5} \exp(34.96 \text{ kJ/mol}/RT) \tag{4}$$

All the characteristics of the α-relaxation of 20 wt.% water in 1-propanol given in the above suggest it has negligible intermolecular cooperativity and zero or nearly zero value for *n*. This can be readily understood if water itself is already highly non-cooperative, and becomes totally non-cooperative in the presence of the much more mobile 1-propanol, which facilitates the α-relaxation of water in the mixture. The prefactor in eq.(4) leads to a frequency that is physically reasonable and consistent with the intermolecular vibrational frequency of water [56, 57]. The small activation energy comparable to the energy to break one to two hydrogen bonds also indicate the α-relaxation of 20 wt.% water in 1-propanol is highly non-cooperative.

*3.4 Water confined in spaces of nanometer size*

Water confined in various spaces with nano-meter in size also relaxes with relaxation times $\tau_{conf}$ having Arrhenius temperature dependence, and as short as the JG β-relaxation time $\tau_\beta$ of water in mixtures with hydrophilic solutes including 2EG, EG, $N_2H_4$ and $H_2O_2$ discussed above. Such dielectric data of $\tau_{conf}$ were found by Hedstrom et al. [58] in water confined in molecular sieves MCM-41 with pore diameter 2.14 nm at hydration levels 12 wt% (open circles), and 22 wt% (open triangles) shown in Fig.4, and previously in Fig.6 in Ref.[30]. Also shown before in Fig.6 in Ref.[30] by its Arrhenius *T*-dependence is the $\tau_{conf}$ of water confined in graphite oxide [59]. Not shown are nearly the same results of $\tau_{conf}$ that were found in water confined in Na-vermiculite clay [60], and in silica gels [61, 62]. Oguni and coworkers have measured the relaxation of water confined within 1.2, 1.6 and 1.8 nm nano-pores of MCM-41 by adiabatic calorimetry, and the fastest relaxation having relaxation time of $10^3$ s at *T*=115 K is shown by the large ⊗ in Fig.4 [63]. The large open square in Fig.4 indicates the value of 1000/*T* at which $\tau_{conf}=10^3$ s was found in water confined within silica pores by adiabatic calorimetry [64]. The fact that $\tau_{conf}$ has properties like $\tau_\beta$ is readily understandable in view of the further reduction of cooperativity and length-scale of the α-relaxation of water when confined in nano-meter space.

We now bring $\tau_{conf}$ into discussion together with the JG β-relaxation time $\tau_\beta$ of water in aqueous mixture with 22.5 and 30.3 mol. % of EG and water solution of hydrazine with 26.5 mol. % of $N_2H_4$. The JG β-relaxation of water in solutions of hydrazine was not resolved in the dielectric spectra, and its relaxation time $\tau_\beta$ is approximated by the primitive relaxation time $\tau_0$ calculated by eq.(2). These JG β-relaxation times are shown in Fig.4 together with the corresponding $\tau_\alpha$. The thick black, red and blue dashed lines in Fig.4 are drawn to suggest plausible temperature dependence of $\tau_\beta$ in the glassy state of the three aqueous mixtures. The red and blue line both ends at $\tau_{conf}$ of water confined in MCM-41 obtained by adiabatic calorimetry [63]. One can observe that $\tau_\beta$ near $T_g$ of the two aqueous mixtures are nearly the same as $\tau_{conf}$ of pure water. This suggests that the dynamics of the two aqueous mixtures are close to that of pure water, and the $T_g$ of pure water is close to 137.8 K and 135.6 K of solution of EG and hydrazine respectively. Structural relaxation time of 33 s for water at 136 K was estimated from the DSC endotherm [23]. Located by the red asterisk in Fig.4, it falls between $\tau_\alpha$ of the two aqueous mixtures. The discussion in the above lends support to the $T_g$=136 K of pure water.

*3.5 Dynamics of water at high temperatures and the purported "fragile" behavior*

The thermodynamic and dynamic properties of bulk liquid water at atmospheric pressure in the temperature range are only accessible to measurements at temperature above the homogeneous nucleation temperature, $T_H$ = 235 K, which is the lower limit of supercooling water. In a paper by Ito et al. [65], it was argued that kinetic 'fragility' of a liquid (*i.e.* degree of deviation from Arrhenius-like dependence in a plot of $\log\eta$ or $\log\tau_\alpha$ vs. $T_g/T$) can be inferred from purely thermodynamic data near and below the melting point. However, they found that water is an exception. Their thermodynamic consideration indicates that near its melting point and for $T > T_H$, water is practically the most 'fragile' of all liquids. On the other hand, near and above the traditionally accepted $T_g$ = 136 K, earlier consideration by Angell [66, 20, 21] and reiterated by Ito et al. has concluded that water is the least fragile. One argument for this is based on the extremely small and broad (relative to $T_g$) increase in the specific heat $C_P$ when glassy water is heated above 136 K, which is a characteristic of a least 'fragile' or a very 'strong' liquid. This ostensibly huge difference in 'fragility' for water at temperatures higher than $T_H$ from that near 136 K, deduced from their own interpretation of thermodynamic behavior, was used by Ito et al. as the basis to propose the existence of a fragile-to-strong transition in supercooled water near 228 K.

This proposal had attracted the attention of others, who looked for the fragile-to-strong transition in confined water [60, 67] and in hydration water of protein [68]. For two layers of water confined between the platelets of sodium vermiculite, dielectric study by Bergman and Swenson [60] found a relaxation with Arrhenius temperature dependence in the range from 125 to 215 K, which they attributed to the α-relaxation of water. They interpreted the observation as support of the proposed 'strong' nature of deeply supercooled bulk water at temperatures below 228 K, and hence the existence of a fragile-strong transition. Invalidity of the interpretation has been pointed out by Johari [69] because the observed dielectric relaxation of the two-molecule thick layer of water contained in sodium vermiculite cannot be identified with the α-relaxation of bulk water. Several reasons were given by Johari, including the temperature dependence of the dielectric strength $\Delta\varepsilon$ of the relaxation which was found to decrease on decreasing the temperature. This contradicts the Curie law of $\Delta\varepsilon$ increasing on decreasing the temperature for α-relaxation, and instead is the signature of secondary relaxation. This criticism of Johari has been accepted by Bergman and Swenson because lately relaxation of water confined in other nanometer spaces with similar nature as in sodium vermiculite has been interpretated as β-relaxation of water by them [58], as well as by others discussed in the previous subsection. The other claimed observation of fragile-strong transition in confined water by neutron scattering [67] was challenged by Swenson [70] and Cerveny et al. [71]. For hydration water of protein, the claim of observing fragile-strong transition found by neutron scattering was dismissed by Doster et al. [72]. They showed the temperature dependence of the relaxation times is highly sensitive to the procedure used in data analysis. While the strong-fragile transition is produced by the use of the standard fitting procedure, it disappears when the same data is analyzed by an improved fitting procedure. From the above, we can conclude so far there is no experimental confirmation of the existence of the strong-fragile transition.

Let us trace back to the physical basis on which Ito et al. concluded from thermodynamic arguments that water is the most fragile of all liquids near the melting point. The difference in entropy between the liquid and crystalline states, $\Delta S$, which decreases with decreasing temperature during supercooling a glass-former was considered by Kauzmann [73]. He plotted the excess entropy $\Delta S$ scaled by its value at the melting point, $\Delta S_m$, against temperature, $T$, normalized by the melting temperature, $T_m$ for six glass-forming liquids, boron trioxide, ethanol, propanol, glycerol, glucose, and lactic acid. This Kauzmann plot allows comparison of the very different manner in which $\Delta S/\Delta S_m$ decreases upon supercooling for these glass-formers. In the Kauzmann plot, glass-formers can be ranked according to the rapidity of their decrease in $\Delta S/\Delta S_m$ as a function of $T/T_m$, or equivalently the proximity of $T_K$ to $T_m$. Ito et al. expanded the original Kauzmann plot by including four additional glass-forming substances: $As_2Se_3$, 3-bromopentane, $Ca(NO_3)_2 \cdot 4H_2O$, and water. The more rapid the drop of $\Delta S/\Delta S_m$ with diminishing $T/T_m$ (and thus larger $T_K/T_m$) in the Kauzmann plot the faster is the decrease of $\log \tau_\alpha$ with lowering of $T_g/T$. This similarity led them to suggest the use of the Kauzmann plot to define ''thermodynamic fragility,'' in analogy with the ''kinetic fragility.'' By this definition, glass-formers having a more rapid drop of $\Delta S/\Delta S_m$ with decreasing $T/T_m$ are more thermodynamically fragile. Ito et al. showed water has the steepest drop of $\Delta S/\Delta S_m$ as a function of $T/T_m$, and hence the highest thermodynamic fragility than most liquids. They made the observation that the decrease of $\Delta S/\Delta S_m$ as a function of $T/T_m$ for the glassformers considered (except water) follows the same order as the decrease of $\log \tau_\alpha$ as a function of either $T/T_g$ or $T_g/T$, the latter gives the Oldekop-Laughlin-Uhlmann-Angell plot used to determine (kinetic) fragility. From the correlation observed between the thermodynamic and kinetic fragilities in the liquids they considered, Ito et al. concluded that the kinetic fragility can be determined purely from thermodynamic data. However, the correlation does not hold in general due to the presence of some glaring exceptions in some well studied glassformers [74, 75, 76]. Since validity of the correlation between kinetic fragility and thermodynamic fragility is neither perfect nor has been established on rigorous theoretical grounds, it is a bold conjecture that water conforms to the correlation, and it is kinetically the most fragile of all liquids based on its thermodynamic fragility. Moreover, as we know, many properties of the dynamics cannot be explained by thermodynamics alone [35].

An independent and direct way to examine whether water at $T > T_H$ is exceptional, as kinetically the most fragile of all liquids, is to compare its viscosity $\eta$ and structural relaxation time [77, 78, 79, 80] at $T > T_H$ with those of other glassformers within approximately the same range of $\eta$ and $\tau_\alpha$. This comparison of $\eta$ of water is presented in Fig.5 with tri-naphthyl benzene (TNB) [81], ortho-terphenyl (OTP) [81], and toluene [81, 82]. Logarithm of the reciprocal of self-diffusion coefficient of water [83] have been shifted vertically to superpose with the viscosity data in the figure. Like the other van der Waals glassformers shown here, and isopropylbenzene (cumene), n-butylbenzene, and n-propyl alcohol with hydrogen bonds [82, 84] not shown here to avoid overcrowding, the temperature dependence of $\eta$ of water is non-Arrhenius at lower temperatures but becomes Arrhenius at higher temperatures, implying that the structural shear relaxation time is also Arrhenius written as $\tau_\eta = \tau_\infty \exp(E_a/RT)$. The value of $\eta$ at infinite temperature obtained by extrapolation of the Arrhenius dependence can be used to determine the prefactor $\tau_\infty$, and the attempt frequency $f_\infty = (1/2\pi\tau_\infty)$. The shear relaxation time $\tau_\eta$ obtained from the Maxwell relation, $\eta(T) = G_\infty(T)\tau_\eta(T)$, where $G_\infty(T)$ is the high frequency shear modulus. For water, ultrasonic measurement have determined $G_\infty(T)/(10^{10}$ dyne/cm$^2) = 1.68 - 0.0127(T - 273)$ [85]. From this we obtain

$$\tau_\eta(T)/\text{s} \approx (2 \times 10^{-15})\exp(14 \text{ kJ/mol}/RT), \tag{5}$$

where the prefactor is equivalent to an attempt frequency of $1.6 \times 10^{14}$ Hz and the activation energy is 14 kJ/mol.

The results for the van der Waals liquids are similar, particularly toluene which has $T_g = 117$ K, and its viscosity is even lower than that of water at the same temperature. From this angle of looking at the dynamics of water, there is nothing exceptional about water to infer that it is kinetically more fragile than all liquids. Shown also in Fig.5 are the viscosity of aqueous solutions of 20.47 and 30.13 mol% of hydrazine [86], and aqueous solutions of ethylene glycol (EG) containing 50 and 75 mol.% of water [87]. The data from the two solutions of hydrazine are just a bit higher that of pure water, and have almost the same temperature dependence. The 50 mol.% water solution of EG is also glassforming. Although higher than bulk water by less than an order of magnitude, the temperature dependence of the viscosity of these mixtures becomes Arrhenius at higher temperature like bulk water, with activation enthalpy slightly higher, which is reasonable because of the presence of the less mobile EG in the solutions.

The Arrhenius $T$-dependence of the relaxation time deduced from $\eta$ of water and given by eq.(5) are about the same as that of the relaxation time obtained from measurement using inelastic X-ray scattering (IXS) [88, 89], Raman scattering [90, 91], and ultrasonic shear measurements [85]. These other sets of data are reproduced from Ref.[88] and shown in Fig.6, where the line is the fit of the IXS data by the Arrhenius $T$-dependence:

$$\tau = (3 \times 10^{-15})\exp(15.9 \text{ kJ}/RT) \text{ s}. \tag{6}$$

There is agreement between eqs.(5) and (6) both on the prefactor and the activation enthalpy.

Dielectric measurements of bulk water at high frequencies above 1 GHz and temperatures higher than $T_H = 235$ K published by several groups are consistent with each other [92, 93, 79, 94]. Their dielectric relaxation times $\tau_D$ of less than $10^{-10}$ s and superposing on each other are presented at the very bottom in Fig.7. The relaxation times from microwave measurements of solution of hydrazine mentioned in a previous section [48, 49] overlap the bulk water data, as well as the relaxation times deduced from diffusion of bulk water data [83, 95]. The lines are VFTH fits to the data of the hydrazine solutions published by Minoguchi et al.[20, 21]. Immediately above are the dielectric relaxation times of solutions of polyvinylalcohol (PVOH), polyvinylpyrrolidone (PVP) [96], and ethylene glycol (EG) with 90, 80, and 40 wt.% of water respectively. The values are close to $\tau_D$ of bulk water. Further above are the relaxation times of 25 wt.% of water nano-confined in graphite oxide obtained by dielectric relaxation [59] and neutron scattering [97, 98]. All these data with relaxation times shorter than 1 ns are enlarged in the inset of Fig.7 for clarity of identification. The temperature dependence of $\tau_D$ at higher temperatures is Arrhenius and is fitted in the inset by the line with

$$\tau_D = 10^{-14}\exp(2009.6 \text{ K}/T) \text{ s}. \qquad (7)$$

The order of magnitude of the prefactor is in agreement and the activation enthalpy (16.7 kJ/mol) is comparable with that determined by viscosity and IXS. Dielectric relaxation measurements of bulk water show it is almost ideally Debye-like with correlation function that is nearly exponential function of time [79, 93]. This behavior is no different from the behavior of many van der Waals liquids such as propylene carbonate [99], and hydrogen bonded liquids such as glycerol and propylene glycol [100]. This is another evidence of showing water is not exceptional in this aspect of the dynamics.

*3.6 Dynamics of aqueous systems at low temperatures*

Dielectric relaxation times longer than 1 ns in Fig.7 are those of 25 wt.% of water nano-confined in graphite oxide, and the α- and β-relaxation times, $\tau_\alpha$ and $\tau_\beta$, of the solutions of EG with 40 wt.% of water and aqueous mixture containing 26.5 and 33 mol.% of hydrazine. Parts of the data of these solutions at lower temperatures have been shown before in Fig.4 for a different purpose. At higher temperatures, $\tau_\beta$ merges with $\tau_\alpha$ at higher temperatures. The relaxation time of water in graphite oxide has Arrhenius temperature dependence over a broad range of temperatures (shown by the dotted brown line in the figure), as expected for the β-relaxation due to nano-confinement. The extrapolation of this Arrhenius $T$-dependence to lower temperatures reaches $10^3$ s very close to the temperature of the fastest relaxation of water confined in molecular sieves found by adiabatic calorimetry also at $10^3$ s [63]. Also this Arrhenius $T$-dependence is close in the value of the relaxation times and the activation energy as the JG β-relaxation of water in aqueous mixture containing 26.5 mol.% of hydrazine that we have constructed before. The diffusion data of water from Smith and Kay [95] have been converted to relaxation times [101], and are also shown in the Fig.7. Actually, the calculation of relaxation time from water self-diffusion coefficient [101] works very well for the high temperature data [83] but its applicability could be doubtful at low temperature. The temperature dependence of data from diffusion in this extremely supercooled region is stronger than that shown by all the solution data and nano-confined water data, although the values are comparable with $\tau_\alpha$ of the two solutions. The diffusion data had previously been used to suggest that water at temperatures between 136 and 150 K is that of an unexceptional 'fragile' liquid with $T_g$ of 136 K [95, 102]. According to us, mixing water with 3EG, 2EG and EG, all having lower mobility than water, should enhance intermolecular interaction and hence 'fragility'. From the fact that all aqueous mixture of 3EG, 2EG and EG are not 'fragile', the diffusion data [95] are doubtful and the accompanying interpretation of water being an unexceptional 'fragile' liquid can be ruled out, same conclusion as made by Minoguchi et al. from a different argument.

Most interesting are the $\tau_\alpha$ of the 40 wt.% water solution of EG (open circles connected by line) measured over 11 decades of time from about 10 s down to $4\times10^{-11}$ s. We can see in either Fig.7 or the inset that the temperature dependence of $\tau_\alpha$ runs parallel to that of $\tau_D$ of bulk water and $\tau_\alpha$ is only about a factor of 4 to 5 longer than $\tau_D$. The $\tau_\alpha$ of 26.5 and 33 mol.% water solutions of hydrazine determined by dielectric measurements with frequencies less than $10^6$ Hz and from microwave measurement had been combined by Minoguchi et al. to yield a global temperature dependence represented by VFTH expression shown by the two lines in Fig.7. Remarkably, the $\tau_\alpha$ from microwave measurements as well as the VFTH fits falls nearly on top of the $\tau_D$ of bulk water. At lower temperatures and in the longer relaxation time region, we have given arguments in previous subsections that the relaxation time of water should be shorter than $\tau_\alpha$ of 40 wt.% water solution of EG, but not shorter than $\tau_\alpha$ of aqueous mixture containing 26.5 mol.% of hydrazine. Combining the relations between the relaxation times of bulk water and the solutions found experimentally at short times and deduced at longer times by arguments, we conclude that the elusive relaxation time $\tau_\alpha$ of water must lie somewhere in between $\tau_\alpha$ of the 40 wt.% water solution of EG and the aqueous mixture containing 26.5 mol.% of hydrazine, had water been prevented from crystallizing below $T_H$. Thus, dynamically the behavior of water is similar to that of the two solutions in temperature dependence of the structural relaxation time, in the degree of cooperativity or non-exponentiality, and in the relation between $\tau_\beta$ and $\tau_\alpha$. Shown also in Fig.7 by the large star is the structural relaxation time ~33 s of ASW water at 136 K, revised by Johari [23] from ~70 s originally estimated from the DSC endotherm [5]. Located within the region bounded by $\tau_\alpha$ of the two solutions, it is consistent with our suggested relaxation time of water. It is worth mentioning here that the study of the dynamics of *dry* oligomers of propylene glycol, monomethyl ether, and dimethyl ether by Mattsson et al. [103]. From the trend of change in the dynamics of these oligomers with number *N* of monomer units in the chain, and visualizing water as the "*N*=0 glycol", they concluded that $T_g$ of water lies within the range, $124 \leq T_g \leq 136$ K. Although their conclusion is similar to ours, their use of *dry* polyethers is totally different in spirit from our consideration of the relation in the dynamic of water with aqueous mixtures with high water content.

The $\tau_\beta$ of water nano-confined in graphite oxide offers another guidance for locating $\tau_\alpha$ of bulk water (see Fig.7) from the fact that $\tau_\alpha$ must be longer than $\tau_\beta$. The measured $\tau_\beta$ of water nano-confined in graphite oxide is not necessarily the $\tau_\beta$ of bulk water because water in the former may have some residual interaction with the walls of the graphite oxide. Nevertheless, the two are not be too different. This can be deduced from the proximity of $\tau_\beta$ of water nano-confined in graphite oxide with the $\tau_\beta$ and $\tau_\alpha$ of the 40 wt.% water solution of EG and aqueous mixture containing 26.5 mol.% of hydrazine.

Despite the fact that the current debate about the location of the glass transition temperature of water is still heated and controversial [24, 25], only few additional evidences can be found in literature inferring its $T_g$ being different from~136 K. The study of thermal desorption at nanoscale in films of ASW by McClure and co-workers [104] were interpreted to indicate a transport mechanism irreconcilable with a bulk diffusion, concluding that "actual self-diffusion coefficient of amorphous water in the temperature range from 150 to 160 K is significantly smaller than previously thought" and that ASW should be in glassy state. On the other hand, these findings are in conflict with those of R. Souda [105] who found from diffusion that a glass-liquid transition of the amorphous water film occurs at 130–145 K. The recent study by Smith and co-workers [106] seems to confirm, by using a similar technique, that water undergoes a glass transition around 136 K. Another recent study by Reinot et al. [107] has reported the absence of rotational motion of the probe Rhodamine 700 (Rh700) embedded in hyperquenched glassy water (HGW) in the temperature range 110 K- 154 K, as revealed by single molecule fluorescence spectroscopy. By comparing this result with those obtained on other liquids (ethanol and OTP), they inferred that viscosity of water should be extremely high in that temperature range and therefore HGW at 154 K is below $T_g$ or its steepness index *m* is extremely small. On the other hand, Leporini [108] disproved the procedure used by Reinot et al. [107] to obtain the viscosity of water from probe rotation, showing that instead all the observations could be compatible with a scenario where water has $T_g$~136 K and a steepness index *m* compatible with that of typical network forming liquids [108]. Moreover, the motion of the larger molecule Rh700 in water is in the hydrodynamic limit, and its relaxation time can be much longer than the structural relaxation time of water. Larger difference between the masses of the host and the guest molecules can lead to larger difference between their relaxation times. Ethanol and OTP are larger molecule than water, and the difference between the structural relaxation time of OTP and Rh700 can be much lesser than in the case of water and Rh700. On the other hand, Reinot et al. clearly stated such as in the caption of their Fig.1: "This schematic is based on our assumption that molecules should rotate on the time scale of minutes when viscosity is close to $10^{12}$ P …". This assumption may not hold, and hence the conclusion made by Reinot et al. could well be invalid.

Therefore, unambiguous literature data, past and present, are consistent with our deduction that the relaxation times of water can be estimated from the various lower and upper bounds obtained from the dynamics of water in mixtures and subjected to nano-confinement as shown in the relaxation map of Fig.7, at all temperatures has it been possible to avoid intervention by crystallization. Accordingly, we conclude from the experimental facts that the glass transition of uncrystallized water occurs at temperature within the range of 130-140 K, similar to the conventional value of $T_g \sim 136$ K. In addition, from the dynamic properties of the various water related systems and bulk water itself, we find that the structural relaxation of water has low degree of cooperativity, and small steepness or 'fragility' index comparable to those of the aqueous mixtures with ethylene glycol and hydrazine presented in Fig.7.

## 4. Conclusion

Taking all the experimental facts and deductions in the previous subsections into consideration, we can conclude that uncrystallized water has low degrees of cooperativity and non-exponentiality (i.e., small $n$), The relation between $\tau_\beta$ and $\tau_\alpha$ of water is similar to that found in the 40 wt.% water solution of EG and aqueous mixture containing 26.5 mol.% of hydrazine, with separation by a few decades at $T_g$ and the two merging together at higher temperatures, consistent with the low degree of non-exponentiality or small coupling parameter, $n$. Its $T_g$ is not far from that of the solution of 26.5 mol.% of hydrazine, and consistent with $\tau_\alpha$ at 136 K of ASW water by calorimetry. The $T_g$-scaled temperature dependence of its $\tau_\alpha$ is expected to be also similar to that of the two solutions ($m$=42 for 40 wt.% water solution of EG), and is characterized by smaller steepness index $m$ or low 'fragility' not too different from that of pure ethylene glycol, and propylene glycol.

Based on their analysis of dielectric data of hydrazine and hydrogen peroxide solutions [20, 21], Minoguchi et al. had given two possible scenarios for the dynamics of water. The first scenario is that water controversially must remain glassy up until the temperature of crystallization which is above 160 K, and thus $T_g$ of water must be higher than 160 K. The second scenario is that water is "an almost ideally strong liquid above 136 K", and thus allowing $T_g$ of water to have the traditional value of 136 K (see also an earlier work with the same conclusion [66]). Actually, this second scenario was considered only in the paper by Minoguchi et al. among many papers published by Angell and coworkers [8, 22, 24, 65] including the latest ones. Notwithstanding, the second scenario is not inconsistent with our deduction of the dynamics of water. All the dynamic properties of water deduced in the section above, and summarized briefly in the previous paragraph are consistent with water behaving like that of a 'strong' liquid throughout the temperature and relaxation time ranges where crystallization can be averted.

From all the dynamic properties of uncrystalized water presented and deductions made in this paper, our answer to the question posed by the title becomes clear and is as follows. The controversy is resolved and $T_g$ of water cannot be anywhere but near the generally accepted value of $T_g \sim 136$ K.

**Acknowledgment**
We thank Ranko Richert for generously sending us the previously published isothermal dielectric loss data of aqueous mixture with 26.5 mol% hydrazine. The data enable us to determine the coupling parameter, and determine the magnitude and temperature dependence of the nearly constant loss and compare that with our data of aqueous mixture EG:H$_2$O in weight proportions of 60:40 and 50:50. Sergiy Ancherbak is warmly thanked for assistance in dielectric spectroscopy measurements.

**Figure Captions**

**Figure 1.** Plot of logarithm of loss factor versus logarithm of frequency for water solution with 26.5 mol % $N_2H_4$ at temperatures from right to left: 162, 160, 158, 156, 154, 142, 150, 146, 144, 142 K. Dashed lines are fit with eq.(3) with n=0.16, 0.16, 0.16, and 0.19 (from right to left at temperature 162 K, 154 K, 150 K, 146 K). Arrows indicate the location of the primitive frequency, $f_0 \equiv 1/(2\pi\tau_0)$, where $\tau_0$ is calculated by eq.(1). Original data have been kindly obtained from R. Richert [20, 21].

**Figure 2.** Plot of loss factor versus temperature for water solution with 26.5 mol % $N_2H_4$ (evaluated at 316 kHz, red circles) and water solution with 22.5 % mol. (50% wt.) ethylene glycol (evaluated at 585 kHz). Inset shows loss spectra of water solution with 22.5% mol. ethylene glycol at selected temperatures: 143, 140, 138, 136, 132, 126, 124, 120, 116, 112, 108, 100, 96 K. Data of 26.5 mol % $N_2H_4$ water solution have been provided by R.Richert.

**Figure 3.** Bilogarithmic plot of normalized loss spectra of water solution with 26.5 mol % $N_2H_4$ (black symbols, at 110 K), glycerol (green symbols, at 179 K), threitol (red symbols, at 224 K), and xylitol (blue symbols at 243 K). Vertical arrows mark the locations of the primitive frequency, $f_0=1/(2\pi\tau_0)$, calculated for each set of data according to Eq.(1). The procedure used to obtain these normalized master data is described in the text. Data of 26.5 mol % $N_2H_4$ water solution have been provided by R. Richert.

**Figure 4.** Comparison of relaxation map for confined and mixed water. Dielectric $\tau_{conf}$ of water confined in molecular sieves MCM-41 with pore diameter 2.14 nm at hydration levels 12 wt% (magenta open circles), and 22 wt% (magenta open triangles) [58], relaxation measured by adiabatic calorimetry of water confined within nano-pores of MCM-41 (large $\otimes$) [63] and water confined within silica pores (large square) [64]. Relaxation time of water in aqueous mixture with 22.5 (red inverted triangles) and 30.3 (black circles) mol. % of EG and water solution of hydrazine with 26.5 mol. % of $N_2H_4$ (blue squares). $\tau_\alpha$ and $\tau_\beta$ are represented by open and filled symbols, respectively. Green line is a VFT fit to $\tau_\alpha$ of the water/hydrazine mixture. Dashed lines suggest the behavior of $\tau_\beta$ in the glassy state for each mixtures. Red asterisk represents the structural relaxation time of water at 136 K as estimated from the DSC endotherm [23].

**Figure 5.** Logarithm of viscosity (or rescaled relaxation time) versus 1000/T for bulk water: red crosses are data from [80], open cyan diamonds are from shifting the logarithm of the reciprocal of self-diffusion coefficient of water [83] and, hidden in the back, open blue diamonds are from [79], green up triangles and yellow asterisks are from [77, 78]. The open blue circles and magenta triangles just above the data of pure water are viscosity for 20.47 and 30.13 mol% of hydrazine respectively [86]. Open red squares and solid magenta triangles are from 50% and 75% mol. of EG in water solution [87]. Open black circles and triangles represent toluene [81, 82], red solid triangles, open green diamonds and solid violet circles are for tri-naphthyl benzene (TNB) [81], black circles are for ortho-terphenyl (OTP) [81]. The lines at high temperatures are linear fits, the low temperature curves are VFT fitting curves.

**Figure 6.** Relaxation map for bulk water at very short times: inelastic X-ray scattering data (filled circles [88] and squares [89]), Raman scattering [90, 91], and ultrasonic shear measurements [85]. Dashed line is eq.(6).

**Figure 7 .** Logarithm of relaxation time versus reciprocal temperature for different systems. At the left bottom dielectric relaxation times $\tau_D$ for bulk water (dark green open square, green open circles and open triangles) from [92, 93, 79, 94] and the relaxation times deduced from diffusion data (cyan solid triangles) [83, 95]; just above (green crossed diamonds) the relaxation times from microwave measurements of solution of hydrazine [48, 49]. The violet and green lines are VFTH fits to the data of the hydrazine solutions published by Minoguchi et al.[20, 21]. Immediately above (blue squares) are the dielectric relaxation times of solutions of polyvinylalcohol (PVOH), polyvinylpyrrolidone (PVP) [96], and ethylene glycol

(EG) with 30 (red open circles), and 40 wt% (black open circles) of water. Relaxation times of 25 wt.% of water nano-confined in graphite oxide obtained by dielectric relaxation [59] and neutron scattering [97, 98] are plotted as brown solid diamonds and dotted brown line is an Arrhenius fit. Adiabatic calorimetry data of water confined within nano-pores of MCM-41 (large ⊗) [63] and water confined within silica pores (large square) [64] are shown. Red asterisk represents structural relaxation time of water at 136 K as estimated from the DSC endotherm [23]. Red open triangles are the diffusion data of water [95] converted to relaxation times. There are reported also the relaxation times of water in aqueous mixture with 50 % wt. (grey inverted triangles) and 40 % wt. (black circles) of EG and water solution of hydrazine with 26.5 mol. % of $N_2H_4$ (blue squares). $\tau_\alpha$ and $\tau_\beta$ are represented by open and filled symbols, respectively. All data with relaxation times shorter than 1 ns are enlarged in the inset. The red line in the inset is an Arrhenius fit of $\tau_D$ at higher temperatures (eq.7).

-------------------

**Figures**

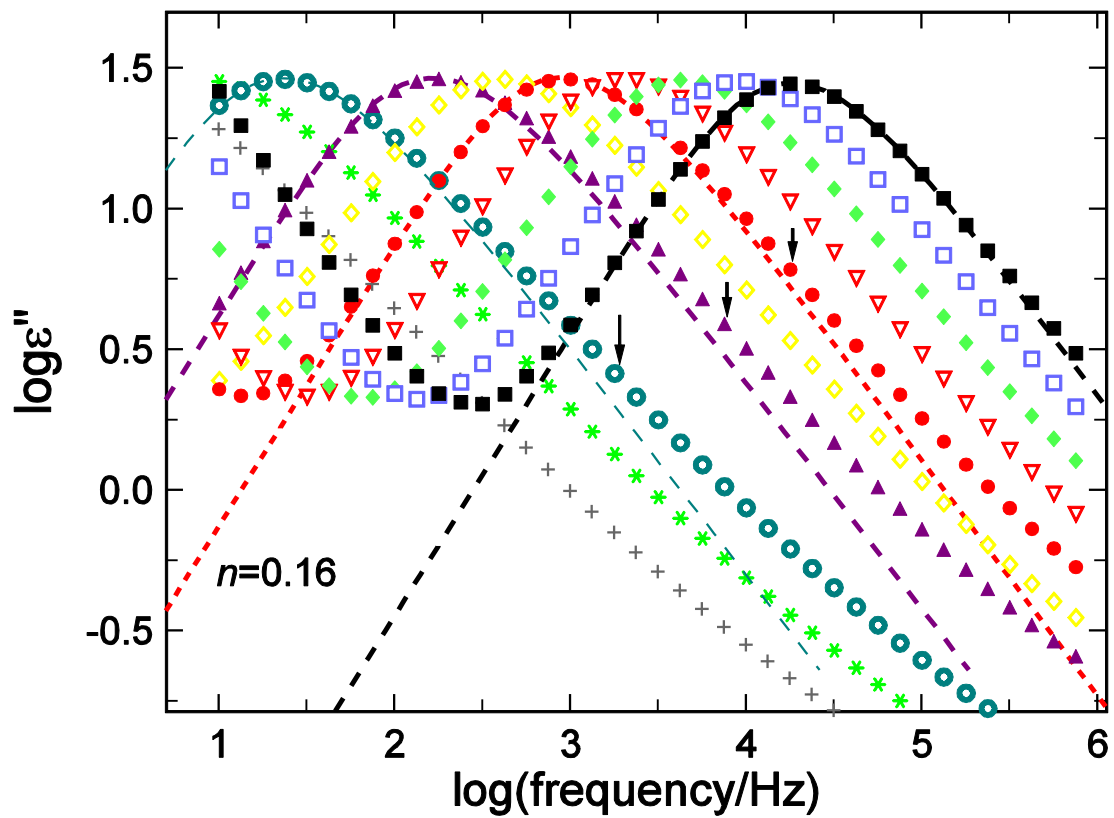

**Figure1**

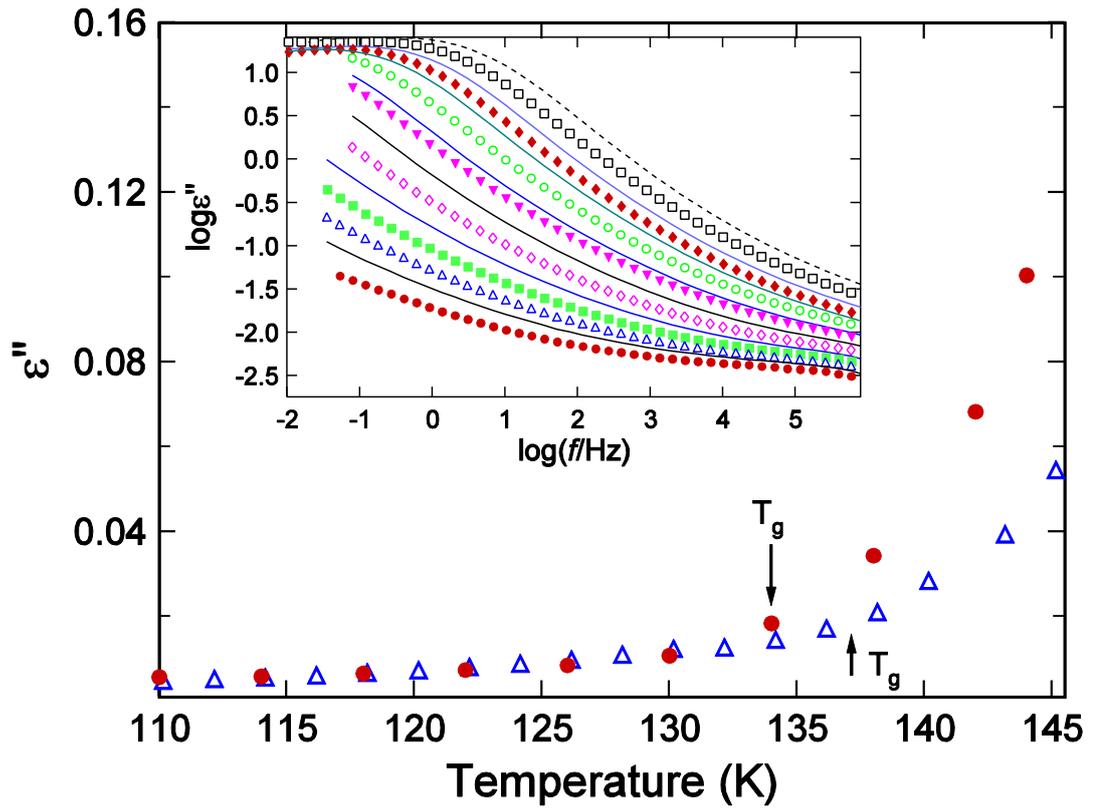

Figure 2

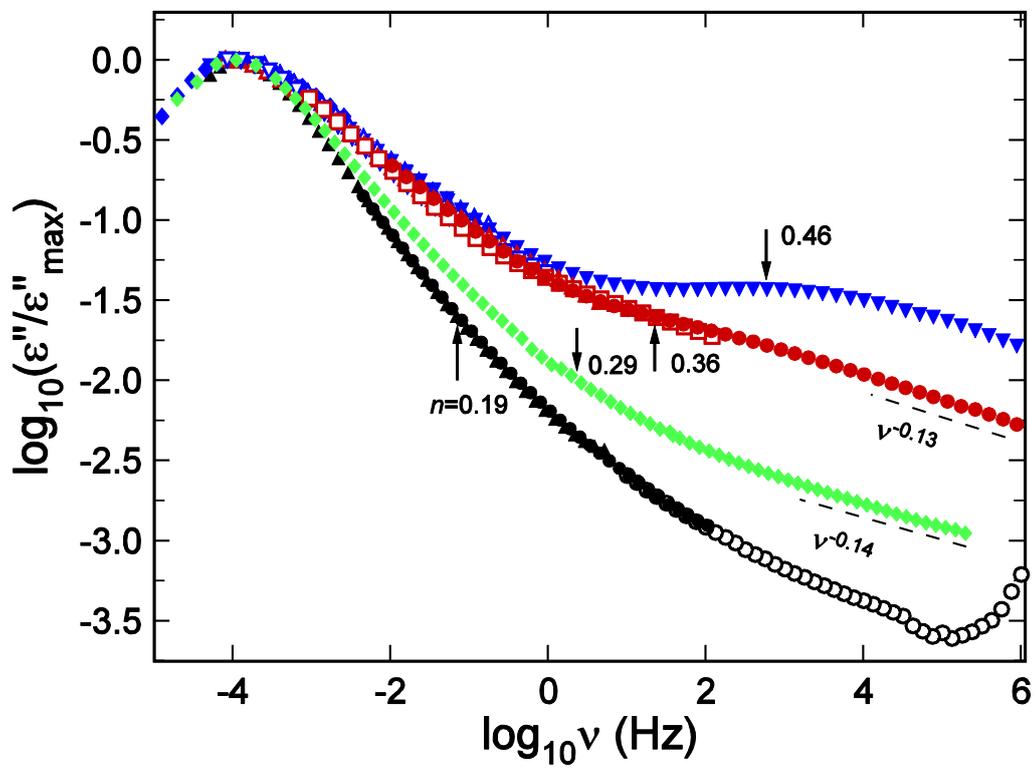

Figure 3

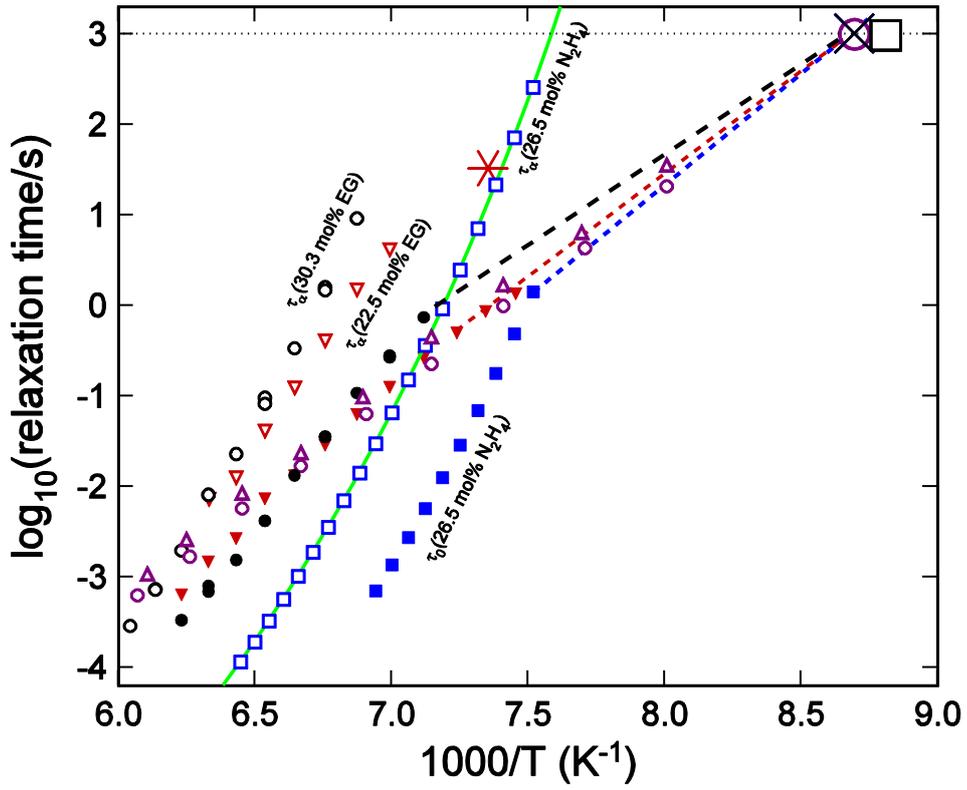

Figure 4

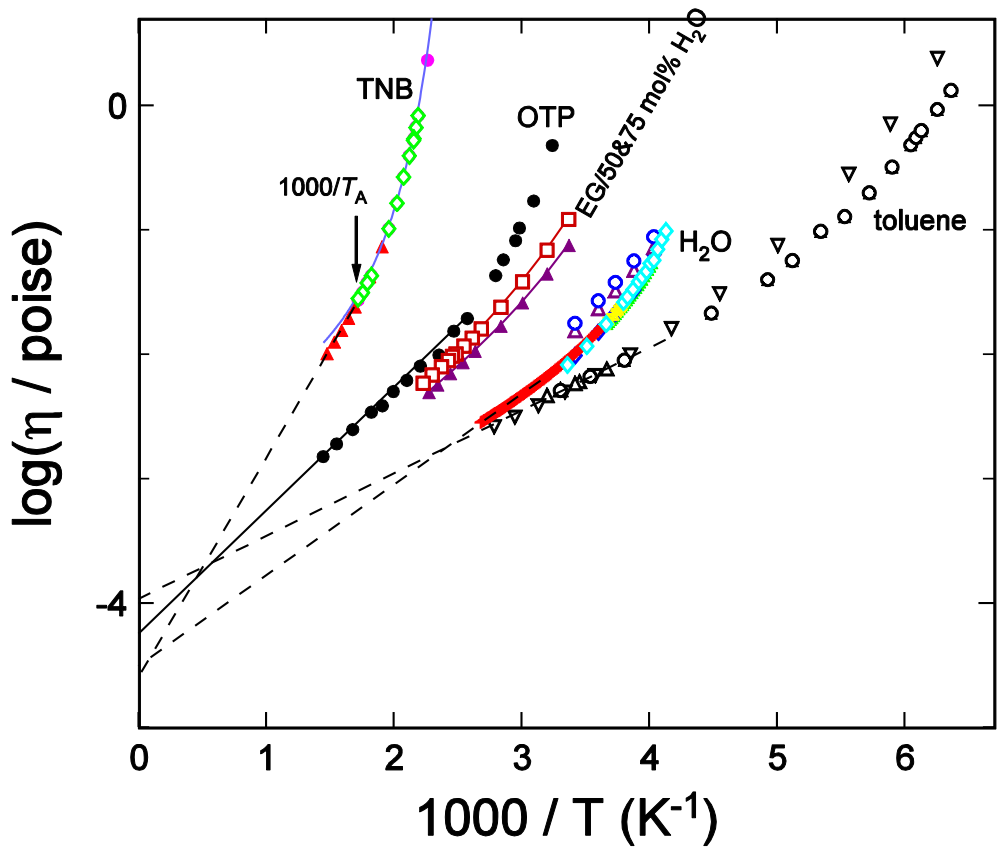

Figure 5

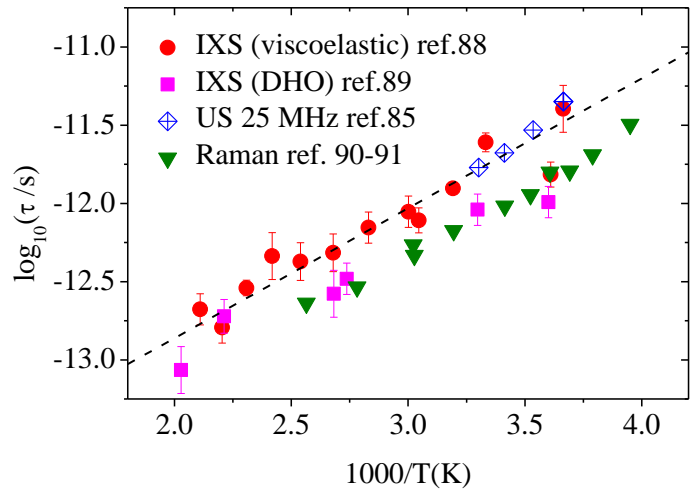

Figure 6

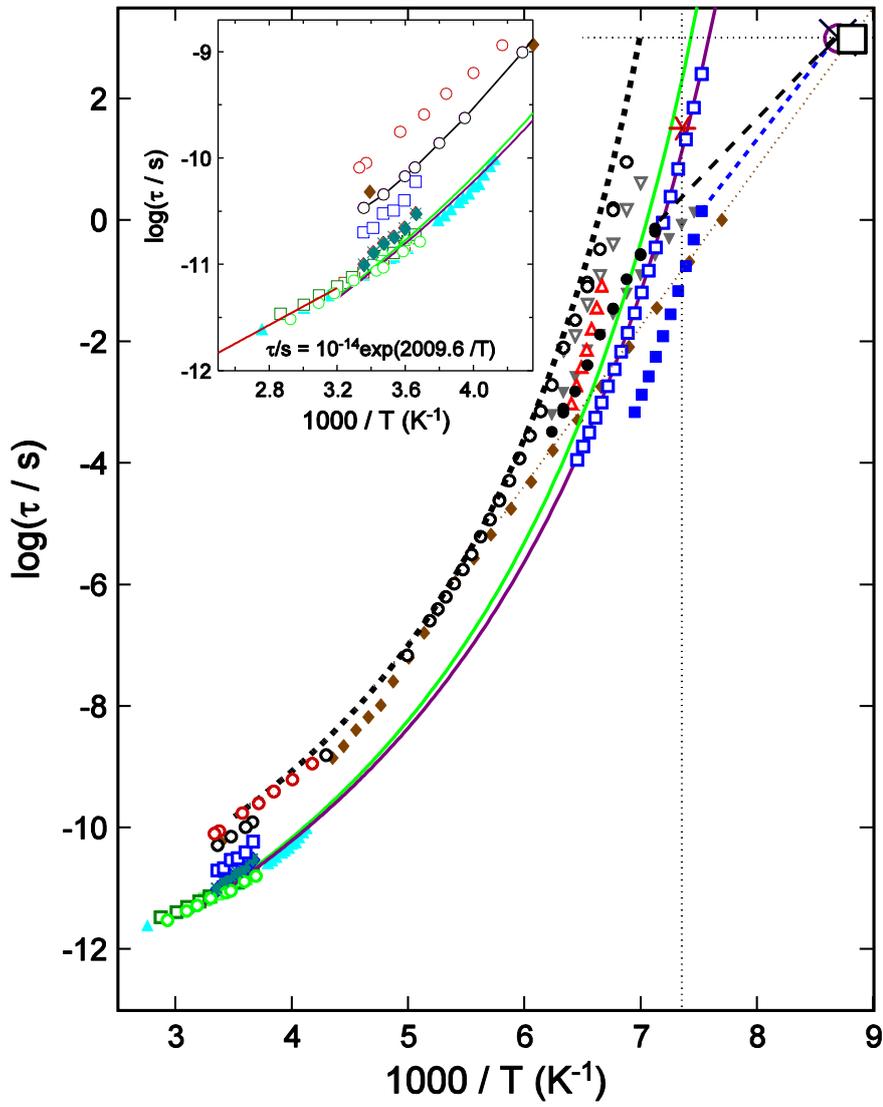

Figure 7

22